\documentclass[10.75pt, a4paper]{article}
\usepackage{titlesec}
\titleformat{\section}
  {\bf\sffamily}
  {\thesection. }
  {5pt}
  {\MakeUppercase}
\renewcommand{\thesection}{\Roman{section}} 

\titleformat{\subsection}
  {\bf\sffamily}
  {\thesubsection. }
  {5pt}{}
  
\renewcommand{\thesubsection}{\Alph{subsection}}

\usepackage[affil-it]{authblk} 
\usepackage{etoolbox}
\usepackage{lmodern}

\usepackage{eurosym}

\usepackage[utf8]{inputenc}
\usepackage{geometry}
\usepackage[hidelinks]{hyperref}
\usepackage[citestyle=numeric,style=phys,biblabel=brackets,backend=bibtex,sorting=none,url=false,doi=false, natbib]{biblatex}
\bibliography{bibliography}
\DefineBibliographyStrings{english}{andothers={\itshape et\addabbrvspace al\adddot}}

\usepackage{csquotes}
\usepackage[]{authblk} 
\usepackage{etoolbox}
\usepackage{lmodern}

\usepackage{amsmath}
\usepackage{enumerate}
\usepackage{enumitem}
\usepackage{graphicx}
\usepackage{siunitx}
\usepackage{float}
\usepackage[]{parskip} 

\makeatletter
\patchcmd{\@maketitle}{\LARGE \@title}{\fontsize{16}{19.2}\selectfont\@title}{}{}
\makeatother

\usepackage[normalem]{ulem}

\usepackage{graphicx}
\usepackage{dcolumn}
\usepackage{bm}
\usepackage{tikz}
\usetikzlibrary{math}
\usetikzlibrary{shapes.geometric, arrows}
\usetikzlibrary{positioning}
\usetikzlibrary{shapes,arrows}
\usetikzlibrary{intersections}
\usepackage{csvsimple}
\usepackage{changepage}
\usepackage[tbtags]{mathtools}
\usepackage{siunitx}
\usepackage{csquotes}

\usepackage{float}
\usepackage{subfigure}
\usepackage[
	nonumberlist, 				
	acronym,      				
	nomain,						
	nopostdot,					
]{glossaries}
\usepackage{glossary-superragged}
\usepackage[hidelinks]{hyperref}
\usepackage{color, colortbl}

\usepackage{wrapfig}

\usepackage{pgfplots}
\usepackage{tikz}
\usetikzlibrary{arrows}
\usepackage{amsmath}
\pgfplotsset{compat=newest}
\usepgfplotslibrary{fillbetween}
\usetikzlibrary{calc}
\def\centerarc[#1](#2)(#3:#4:#5)
{ \draw[#1] ($(#2)+({#5*cos(#3)},{#5*sin(#3)})$) arc (#3:#4:#5); }

\usepackage{amssymb}

\usepackage{nicefrac}

\usepackage{abstract}

\usepackage{multirow}
\usepackage{tabularx}
\usepackage{booktabs}
\usepackage{array}
\usepackage{longtable}
\newcolumntype{L}[1]{>{\raggedright\let\newline\\\arraybackslash\hspace{0pt}}m{#1}}
\newcolumntype{C}[1]{>{\centering\let\newline\\\arraybackslash\hspace{0pt}}m{#1}}
\newcolumntype{R}[1]{>{\raggedleft\let\newline\\\arraybackslash\hspace{0pt}}m{#1}}

\newacronym{3d}{3D}{three dimensional}
\newacronym{am}{AM}{additive manufacturing}
\newacronym{fdm}{FDM}{fused deposition modeling}
\newacronym{ism}{ISM}{in-space manufacturing}
\newacronym{iss}{ISS}{International Space Station}
\newacronym{fcb}{FCB}{Functional Cargo Block}
\newacronym{dem}{DEM}{discrete element method}
\newacronym{md}{MD}{molecular dynamics}
\newacronym{dc}{DC}{direct-current}
\newacronym[plural=PFCs,firstplural=parabolic flight campaigns (PFCs)]{pfc}{PFC}{Parabolic Flight Campaign}
\newacronym{fft}{FFT}{Fast Fourrier Transform}
\newacronym{cad}{CAD}{Computer Assisted Design}
\newacronym{ptfe}{PTFE}{polytetrafluoroethylene}
\newacronym{ps}{PS}{polystyrene}
\newacronym{nasa}{NASA}{National Aeronautics and Space Administration}
\newacronym{esamm}{ESAMM}{Extended Structure Additive Manufacturing Machine}
\newacronym{amf}{AMF}{Additive Manufacturing Facility}
\newacronym{us}{US}{United States}
\newacronym{usa}{USA}{United States of America}
\newacronym{bmgs}{BMGs}{Bulk Metallic Glasses}
\newacronym{esa}{ESA}{European Space Agency}
\newacronym{si}{SI}{International System of Units, abbreviated from French \textit{Syst\`{e}me International (d'unit\'{e}s)}}
\newacronym{dlr}{DLR}{German Aerospace Center}
\newacronym{liggghts}{LIGGGHTS}{\acrshort{lammps} Improved for General Granular and Granular Heat Transfer Simulations}
\newacronym{lammps}{LAMMPS}{Large-scale Atomic/Molecular Massively Parallel Simulator}
\newacronym{sjkr}{SJKR}{Simplified Johnson-Kendall-Roberts}
\newacronym{ded}{DED}{Directed Energy Deposition}
\newacronym{slm}{SLM}{Selective Laser Melting}
\newacronym{sls}{SLS}{Selective Laser Sintering}
\newacronym{eva}{EVA}{Extra-Vehicular Activity}
\newacronym{sem}{SEM}{Scanning Electron Microscopy}
\newacronym{RPM}{RPM}{Ramdom Positioning Machine}
\newacronym{rpm}{rpm}{revolutions per minute}
\newacronym{rise}{RISE}{Research Internships in Science and Engineering}
\newacronym{daad}{DAAD}{German Academic Exchange Service, abbreviated from German \textit{Deutscher Akademischer Austauschdienst}}
\newacronym{fsm}{FSM}{finite-state machine}
\newacronym{ir}{IR}{infrared}
\newacronym{pcbs}{PCBs}{Printed Circuit Boards}
\newacronym{pcb}{PCB}{Printed Circuit Board}
\newacronym{mcr}{MCR}{Modular Compact Rheometer}
\newacronym{sff}{SFF}{Solid Freeform Fabrication}
\newacronym{uv}{UV}{ultraviolet}
\newacronym{abs}{ABS}{acrylonitrile butadiene styrene}
\newacronym{hpde}{HPDE}{high density polyethylene}
\newacronym{pei}{PEI}{polyetherimide}
\newacronym{bff}{BFF}{BioFabrication Facility}
\newacronym{lens}{LENS}{Laser Engineered Net Shaping}
\newacronym{cnc}{CNC}{Computer Numerical Control}
\newacronym{ebf3}{EBF$^3$}{Electron Beam Free-Form Fabrication}
\newacronym{leo}{LEO}{Low Earth Orbit}
\newacronym{pc}{PC}{polycarbonate}
\newacronym{crissp}{CRISSP}{Customisable Recyclable International Space Station Packaging}
\newacronym{Athena}{Athena}{Advanced Telescope for High-ENergy Astrophysics}
\newacronym{lbm}{LBM}{Laser Beam Melting}
\newacronym{bam}{BAM}{Federal Institute for Materials Research and Testing, abbreviated from German \textit{Bundesanstalt f\"{u}r Materialforschung und-pr\"{u}fung}}
\newacronym{pbf}{PBF}{powder bed fusion}
\newacronym{eb}{EB}{Electron Beam}
\newacronym{2d}{2D}{two dimensional}
\newacronym{4d}{4D}{four dimensional}
\newacronym{ft4}{FT4}{Freeman Technology 4 Powder Rheometer}
\newacronym{dsc}{DSC}{Differential Scanning Calorimetry}
\newacronym{pmma}{PMMA}{polymethylmethacrylate}
\newacronym{1g}{$1g$}{gravity on-ground}
\newacronym{mug}{$\mu g$}{microgravity}
\newacronym{bcm}{BCM}{Box Counting Method}
\newacronym{mct}{MCT}{Mode Coupling Theory}
\newacronym{gmct}{gMCT}{granular Mode Coupling Theory}
\newacronym{itt}{ITT}{Integration Through Transients}
\newacronym{mfc}{MFC}{Mass Flow Controller}
\newacronym{ct}{CT}{computed tomography}
\newacronym{xct}{XCT}{X-ray computed tomography}
\newacronym{cv}{CV}{curriculum vitae}
\newacronym{pi}{PI}{principal investigator}
\newacronym{osp}{OSP}{orthogonal superimposed perturbation}
\newacronym{npi}{NPI}{Network Partnering Initiative}
\newacronym{ecsat}{ECSAT}{European Centre for Space Applications and Telecommunications}
\newacronym{eac}{EAC}{European Astronaut Centre}
\newacronym{estec}{ESTEC}{European Space Research and Technology Centre}
\newacronym{fps}{fps}{frames per second}
\newacronym{pdf}{pdf}{probability density function}
\newacronym{al}{Al}{aluminium}
\newacronym{ss}{\textit{SS}}{\textit{Smooth Surface}}
\newacronym{rs}{\textit{RS}}{\textit{Rough Surface}}
\newacronym{rcp}{rcp}{random close packing}
\newacronym{iop}{IoP UvA}{Institute of Physics of the University of Amsterdam}
\newacronym{mp}{MP}{Institute of Material Physics for Space}
\newacronym{elgra}{ELGRA}{European Low Gravity Research Association}
\newacronym{zarm}{ZARM}{Center of Applied Space Technology and Microgravity}
\newacronym{piv}{PIV}{particle image velocimetry}
\usepackage[framemethod=tikz]{mdframed}
\usepackage{lipsum}
\usepackage{dirtytalk}
\usepackage{tcolorbox}
\usepackage[font=footnotesize]{caption}
\newtcolorbox{mybox}[1]{colback=green!6!white,colframe=black!75!black,fonttitle=\bfseries,title=#1}
\newtcolorbox{mybox2}{colback=red!5!white,colframe=red!75!black}

\usepackage{pifont}

\usepackage{soul,xcolor}
\setstcolor{red}

\usepackage{xcolor,hyperref}
\hypersetup{
   colorlinks,
   linkcolor={blue!50!black},
   citecolor={blue!50!black},
   urlcolor={blue!80!black}
} 

\definecolor{mycolor}{rgb}{0.122, 0.435, 0.698}

\title{Viscoplastic Lines: \\
Printing a Single Filament of Yield Stress Material on a Surface}
 \author[1]{Jesse van der Klok}
  \author[1]{Daniël Tieman}
   \author[1]{Maziyar Jalaal\footnote{m.jalaal@uva.nl}}

\affil[1]{Van der Waals-Zeeman Institute, Institute of Physics, University of Amsterdam,\\

Science Park 904, Amsterdam, 1098XH, The Netherlands}

\begin{document}
\definecolor{brickred}{rgb}{0.8, 0.25, 0.33}
\definecolor{darkorange}{rgb}{1.0, 0.55, 0.0}
\definecolor{persiangreen}{rgb}{0.0, 0.65, 0.58}
\definecolor{persianindigo}{rgb}{0.2, 0.07, 0.48}
\definecolor{cadet}{rgb}{0.33, 0.41, 0.47}
\definecolor{turquoisegreen}{rgb}{0.63, 0.84, 0.71}
\definecolor{sandybrown}{rgb}{0.96, 0.64, 0.38}
\definecolor{blueblue}{rgb}{0.0, 0.2, 0.6}
\definecolor{ballblue}{rgb}{0.13, 0.67, 0.8}
\definecolor{greengreen}{rgb}{0.0, 0.5, 0.0}
\begingroup
\sffamily
\date{}
\maketitle
\endgroup

\begin{abstract}

This study presents the spreading of a single filament of a yield stress (viscoplastic) fluid extruded onto a pre-wetted solid surface. The filaments spread laterally under surface tension forces until they reach a final equilibrium shape when the yield stress dominates. We use a simple experimental setup to \emph{print} the filaments on a moving surface and measure their final width using optical coherence tomography. 
Additionally, we present a scaling law for the final width and determine the corresponding pre-factor using asymptotic analysis. {\color{black}We then analyse the level of agreement between the theory and experiments and discuss the possible origins of discrepancies.} The process studied here has applications in extrusion-based thermoplastic and bio-3D printing.\\


\textbf{keywords: Yield Stress $|$ Viscoplastic Fluids $|$ Plastocapillarity $|$ Surface Tension $|$ 3D Printing} 

\end{abstract}


\section{INTRODUCTION}
Spreading of non-Newtonian liquids on surfaces has widespread interest in many fabrication, coating, and printing applications~\citep{derby2010inkjet, lohse2022fundamental, mackay2018importance}. Among them are the extrusion-based three-dimensional (3D) printing techniques, where digitally-designed objects are built via line-by-line and then layer-by-layer deposition of the ink through a nozzle. A diverse range of materials can be used in 3D printing for various manufacturing purposes, from soft tissues and synthetic bones in biomedical applications~\citep{placone2018recent} to food pastes~\citep{zhu2019extrusion}, {\color{black}energy materials}~\citep{tagliaferri2021direct}, and ceramics~\citep{faes2015extrusion}. 
{\color{black}These materials, depending on the application, are printed as single filament or monolayer (\textit{e.g.}, for conductive circuit printing \citep{gnanasekaran20173d, postiglione2015conductive, valino2019advances, zhang2016fabrication}) or, more commonly, with multi-layer printing \citep{lewis2006direct,ngo2018additive}. The characteristics of the single line depositions can either directly influence the product quality for single filament printing and can also affect the overall quality of multilayer printing via setting up the first layer deposition.}
%
Hence, understanding the underlying physics of deposition of a single line of inks and the effects of rheological properties is essential to improve the design of extrusion-based 3D printers or to design new inks based on the rheology of complex fluids~\citep{ewoldt2021designing}.
We will address the deposition of a single filament of viscoplastic (or yield stress) fluids. Such materials are neither solids nor liquids. When the external stress is below the yield stress, the material effectively behaves like an elastic solid, but when the stress exceeds the yield stress, the material plastically deforms and flows like a viscous liquid~\citep{barnes1999yield,coussot2014yield,balmforth2014yielding,bonn2017yield}. Viscoplasticity is a common feature of many industrial fluids, such as waxy oil and cosmetic and dairy products. 
{\color{black}Furthermore, viscoplastic fluids represent a large group of polymeric inks used in 3D printing~\citep{lewis2006direct, o2017three}, where, depending on the application, the values of the ink's yield stress varies~\citep{del2021rheological, kiyotake2019development, eom2019rheological, siacor2021additive, siqueira2017cellulose, paxton2017proposal, m2017linking}. While many inks feature a relatively large yield stress of hundreds of Pascals, many others are much softer with yield stress of tens of Pascals. Examples of the latter category can be found in conductive inks and energy materials used in Direct Writing (DW)~\citep{zhang2016fabrication, tagliaferri2021direct}, bioprinting hydrogels~\citep{rodell2013rational,mouser2016yield, wilson2017shear}, Kaolin-based materials~\citep{sun2018creation}, colloidal inks~\cite{eom2019rheological}, and cellulose suspension~\citep{jiang2020rheology}. The main focus of the present work is to study the spreading of such soft viscoplastic inks, where the values of yield stress is comparable with the driving spreading stresses.}

{\color{black}Spreading and deposition of viscoplastic fluids share similarities and also present different characteristics in comparison to Newtonian fluids~\citep{bonn2009wetting}. Like Newtonian (and viscoelastic) fluids, viscoplastic fluids (if soft enough) spreads due to surface tension and gravity, on time scales associated with the apparent viscosity.} The most distinctive trait of viscoplastic spreading, however, might manifest in the final shape. {\color{black}Theoretically}, a Newtonian free-surface flow on a fully-wetting surface spreads (due to gravity or surface tension) until it is finally completely flat~\citep{tanner1979spreading, tuck1990numerical, backholm2014capillary, bergemann2018viscous, jalaal2019capillary}. In contrast, in the same situation, a viscoplastic fluid eventually stops at a finite time, when the stress everywhere inside the material falls below the yield stress. Hence, contrary to a Newtonian liquid, a viscoplastic liquid reaches a characteristic final shape~\citep{dubash2009final, liu2016two, jalaal2021spreading} on a fully-wetting surface. Finding the final shape for spreading viscoplastic filaments is indeed the main objective of this study. {\color{black} Note that, in practice, it might be difficult to distinguish the deposition behaviour of a viscoplastic fluids from a highly viscous Newtonian fluid or a viscoelastic liquid with long characteristic elastic time scales. Moreover, the final shape of the complex fluid could be affected by other factors such as time-dependent material properties like thixotropy~\citep{uppal2017dynamics,oishi2019normal,sen2021thixotropy} and solidification~\citep{tavakoli2014spreading, jalaal2018gel, koldeweij2021initial}.}

The present study relates to a number of previous research lines in the literature. One is the gravity-driven spreading and large-scale 3D printing of yield stress materials, in particular concrete~\citep{roussel2005fifty, flatt2006linking, ancey2007plasticity, liu2016two, buswell20183d, gosselin2016large, valette2021viscoplastic, garg2021fluidisation}. The major difference between the present study and these works is the characteristic lengths scales and the driving mechanisms. While gravity is the main force of spreading in large-scale 3D printers, surface tension is the dominant factor at a small scale or when gravity is absent~\citep{brutin2009sessile,diana2012sessile,d2022spreading}. 
This changes the underlying physics, where the competition of yield stress and capillary forces determines the final shape of the liquid. We refer to the phenomena of capillary-driven yield stress fluids as plastocapillarity. Also related to the present study is the previous investigations on the impact and spreading of viscoplastic droplets on dry and wet surfaces~\citep{luu2009drop, saidi2010influence, german2010spreading, jalaal2015slip, blackwell2015sticking, oishi2019impact, oishi2019normal, sen2021thixotropy, martouzet2021dynamic}, where depending on the regime parameters, different factors such as inertia, elasticity, gravity, {\color{black} thixotropy}, and surface tension can determine the dynamics of the droplets. In the present work, we will focus on a different geometry (filaments) on a small scale and a {\color{black} regime and type of material where surface tension and yield stress effects dominate. }

To address the problem, we will use experimental and theoretical tools. The article is organized as follows: Section~\ref{sec:experiments} presents the experimental details, including the setup and the used fluids. Section~\ref{sec:analys} provides the results, a simple scaling law and a more detailed description of the empirical observations based on viscoplastic lubrication theory. Section~\ref{sec:conc} concludes the results and present future perspectives.

\section{Experiments}
\label{sec:experiments}

\subsection{Setup}

\begin{figure}
  \centerline{\includegraphics[width=14cm]{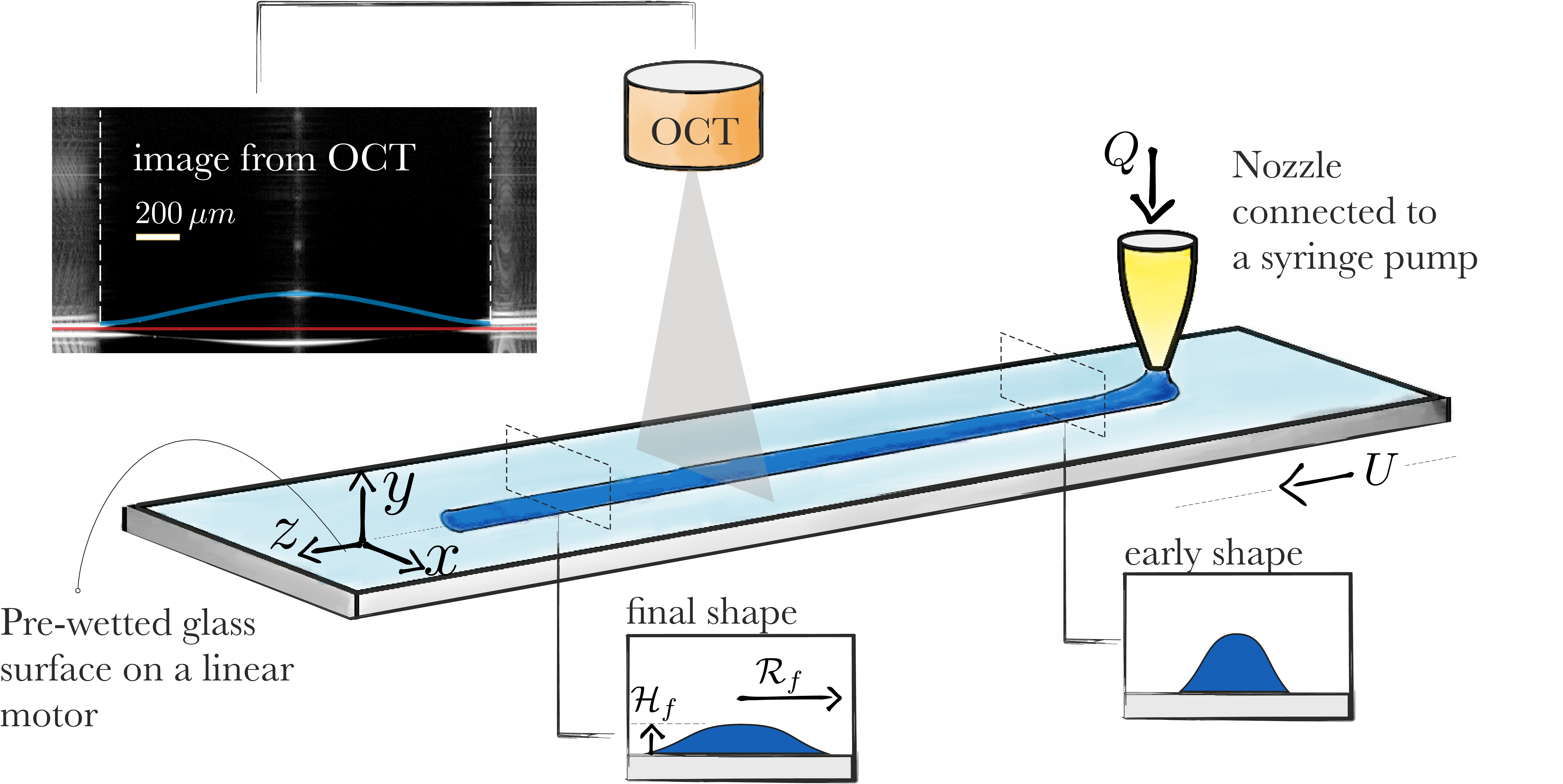}}
  \caption{Experimental setup for the investigation of printing a single line of viscoplastic fluids. The shown dimensions in the schematic are not to scale. A filament of a yield stress (viscoplastic) fluid is printed on a moving pre-wetted substrate. After extrusion from the nozzle, the filament spreads on the surface until it reaches a final width. We note the final half-width of the filament with $\mathcal{R}_f$. An Optical Coherence Tomographer (OCT) is used to measure the final width of the filaments. An example of the OCT image is shown in the top-left of this figure, where blue and red lines highlight the interface of the filament and the substrate. The dashed vertical lines show the edge of the filament, where the surface becomes flat.}
\label{fig:setup}
\end{figure}

The experimental setup is a \say{simplified 3D printer} (see figure~\ref{fig:setup}). The fluid was injected, using a syringe pump, at a constant rate $Q$ through a hydrophobic nozzle with an inner diameter of 0.912\,mm. The substrate was a glass plate, pre-wetted with a thin layer of the same fluid at a constant height of $h_{\infty} \approx 40\,\mu$m. The pre-wetting layer not only simulates the condition of a fully wetted surface but also allows us to avoid any complexity due to the (triple) contact line~\citep{oron1997long, craster2009dynamics, jalaal2021spreading,martouzet2021dynamic}. Pre-wetting 
was done just before the experiments to minimize evaporation effects. The distance between the nozzle and the surface of the pre-wetted film was fixed at $350\pm10\, \mu$m. {\color{black} Changing this distance between $75-400\, \mu$m did not change the results.} The substrate was placed on a linear motor (Thorlabs LTS300/M). When the flow is established through the nozzle, the linear motor moves (constant velocity of $U=$10\,mm$/$s and initial acceleration of 50\,mm$/$s$^2$) and a viscoplastic line begins to form on the surface. The substrate motion continued until a line of length $80\pm 2\,$mm was obtained. Note that the short delays between the pump's function and the substrate's motion disturb the filament's shape at the beginning and end ($\sim 5$mm on each side). We perform the measurements far from these areas. Also, measurements were performed 20s after the deposition to ensure that the final equilibrium shape was obtained.
As the line is printed, it travels under a lens of an Optical Coherence Tomography (OCT) scanner to obtain the shape of the interface.  OCT is a non-invasive optical technique that uses low-coherence interferometry to obtain depth scans~\citep{fercher2003optical}.  A series of these depth scans are then used to form a cross-sectional image. The technique has been used to study evaporating droplets~\citep{edwards2018density}, flow inside channels~\citep{daneshi2019characterising}, and spreading and solidification of droplets of complex fluids~\citep{jalaal2018gel}. The current experimental setup did not allow for an accurate measurement of dynamic of spreading. Nonetheless, using OCT provides many advantages for our system. In the setup, the flat pre-wetted film, unlike the curved surface of the printed line, forms high-intensity patterns of vertical lines. This is due to the over-saturation of the reflected light on a flat surface. Hence, by detecting the edge of these patterns (vertical dashed lines in figure \ref{fig:setup}), we were able to easily obtain the width of the filament with an accuracy of $\mathcal{O}(10\, \mu$m$)$. Furthermore, by adding microparticles (3\,$\mu$m polystyrene from microparticles GmbH) to the fluid, we were able to roughly inspect
the flow inside the printed lines
and
ensure that the flow along the length of the filament (z-direction in figure \ref{fig:setup}) was negligible, and therefore the spreading was mainly across the width (x-y plane in figure \ref{fig:setup}). Experiments were repeated at least 10 times for each data point (total of $\sim$300 experiments). All measurements were performed at steady-state when the filaments had reached their final shape. For each experiment, the OCT scans were performed for four cross-sections of the line, and the values were averaged.

\subsection{Fluids}

\begin{figure}
  \centerline{\includegraphics[width=12cm]{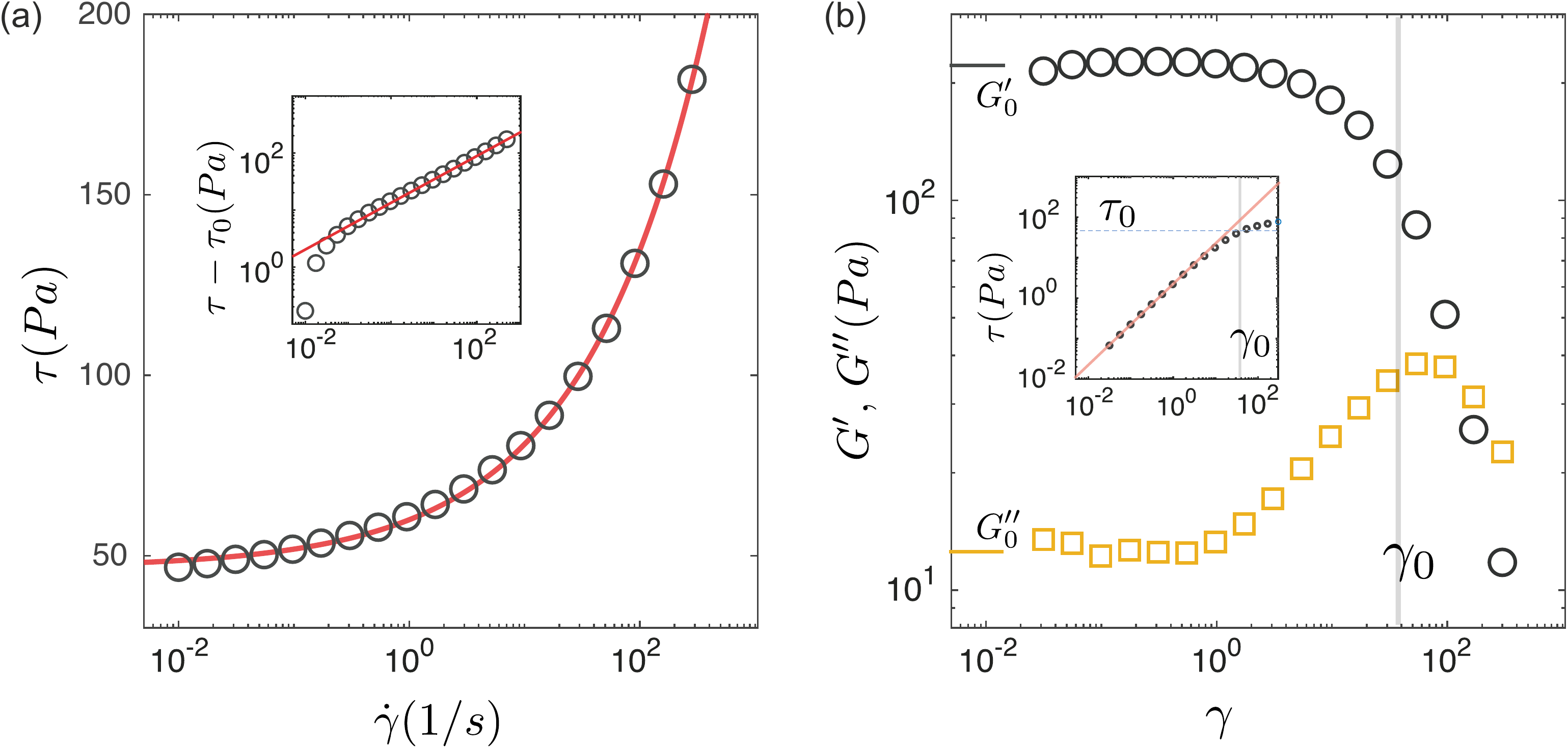}}
  \caption{a) Flow curve (shear stress versus shear rate) for sample 4 (see Table \ref{tab:rheology}). The inset shows the viscous stress ($\tau-\tau_0$) versus the shear rate. The symbols are the measurements, and the red lines are the corresponding Herschel–Bulkley fits. b) Storage modulus ($G^{\prime}$, circles) and loss modulus ($G^{\prime \prime}$, squares) for sample 3 as a function of strain. The vertical grey line shows the critical strain at which stress is equal to the yield stress (from the Herschel-Bulkley fit). The inset shows the variation of shear stress versus the strain. The red line is a linear fit with the slope $G^{\prime}_0 \approx 223$. The normal lines highlight the critical strain and the yield stress.}
\label{fig:rheol}
\end{figure}

We used a mixture of Mili-Q water and a commercial hair gel at five different concentrations. The final materials are basically aqueous mixtures of Carbopol that are pH-neutralized with triethanolamine~\citep{dinkgreve2016different}. The samples were centrifuged before the experiments to remove bubbles. 
An Anton Paar MCR 302 rheometer with a rough cone-and-plate configuration ($2^{\circ}$ angle) was used to characterize the rheological properties of the liquids. For each sample, we obtained the steady-state flow curves as well as viscoelastic response in an oscillatory shear test. Figure~\ref{fig:rheol} shows an example of measurements for one of the samples.
Similar to what has previously been reported~\citep{kim2003rheological,dinkgreve2016different, jalaal2019laser}, the fluids show a Herschel-Bulkley-type behaviour where the viscosities drop with the shear rate, and at low shear rates, the stress values converge to the yield stress. We therefore characterize the material properties by fitting the following equations to the flow curves (see figure~\ref{fig:rheol}a):
\begin{equation}
\tau - \tau_0 = K \dot{\gamma}^n
\end{equation}
where $\tau_0$ is the yield stress, $K$ is the consistency index, and $n$ is the shear index. The fitting values for {\color{black}6 samples} are listed in Table~\ref{tab:rheology}.
{\color{black}Oscillatory tests reveal the elastic-dominated behaviour at low strains/stresses with almost constant elastic and loss moduli.
We quantifies these values by averaging the values of $G^{\prime}$ and $G^{\prime \prime}$ at small strain region, when $\gamma <1$ (see Table~\ref{tab:rheology}).} At large deformations, when stress exceeds the yield stress, the values of $G^{\prime}$, and $G^{\prime \prime}$ drop significantly, and the material mostly behaves like a viscous fluid with weak elasticity (also see~\cite{donley2020elucidating}).

\begin{table}
  \begin{center}
\def~{\hphantom{0}}
  \begin{tabular}{cccccc}
     Sample  & $\tau_0$ (Pa)   &   $K$(Pa\,s$^n$) & $n$ &   $G^{\prime}_0$ (Pa) & $G^{\prime\prime}_0$ (Pa) \\[3pt]
       ~~1~   & 102.0 & ~~34.3~ & 0.39 & ~~365~ & 22.9 \\\
       ~~2~   & 84.2 & ~~27.9~ & 0.40 & ~~327~ & 20.4 \\\
       ~~3~   & 66.2 & ~~21.1~ & 0.40 & ~~282~ & 19.6 \\\
       ~~4~   & 46.6 & ~~13.33~ & 0.41 & ~~223~ & 13.2 \\\
       ~~5~   & 29.8 & ~~8.57~ & 0.42 & ~~149~ & 8.8 \\\
       ~~6~   & 4.0 & ~~1.46~ & 0.51 & ~~25.1~ & 1.7 \\\
  \end{tabular}
  \caption{Values of yield stress $\tau_0$, consistency index $K$, shear index $n$, and small strain elastic module $G^{\prime}_0$, and loss module $G^{\prime \prime}_0$.}
  \label{tab:rheology}
  \end{center}
\end{table}

The density of the fluids ($\rho$) are almost the same as water~\citep{jalaal2019laser}. Like other yield stress fluids, the surface tension of the materials ($\sigma$) is hard to measure. Previous attempts in the literature~\citep{jorgensen2015yield} mostly report slightly smaller values than the surface tension of water. We therefore use $\sigma = 0.072$ N/m in our calculations, but one should take this value likely as an upper bound.  

\section{Analysis}
\label{sec:analys}
\begin{figure}
  \centerline{\includegraphics[width=12cm]{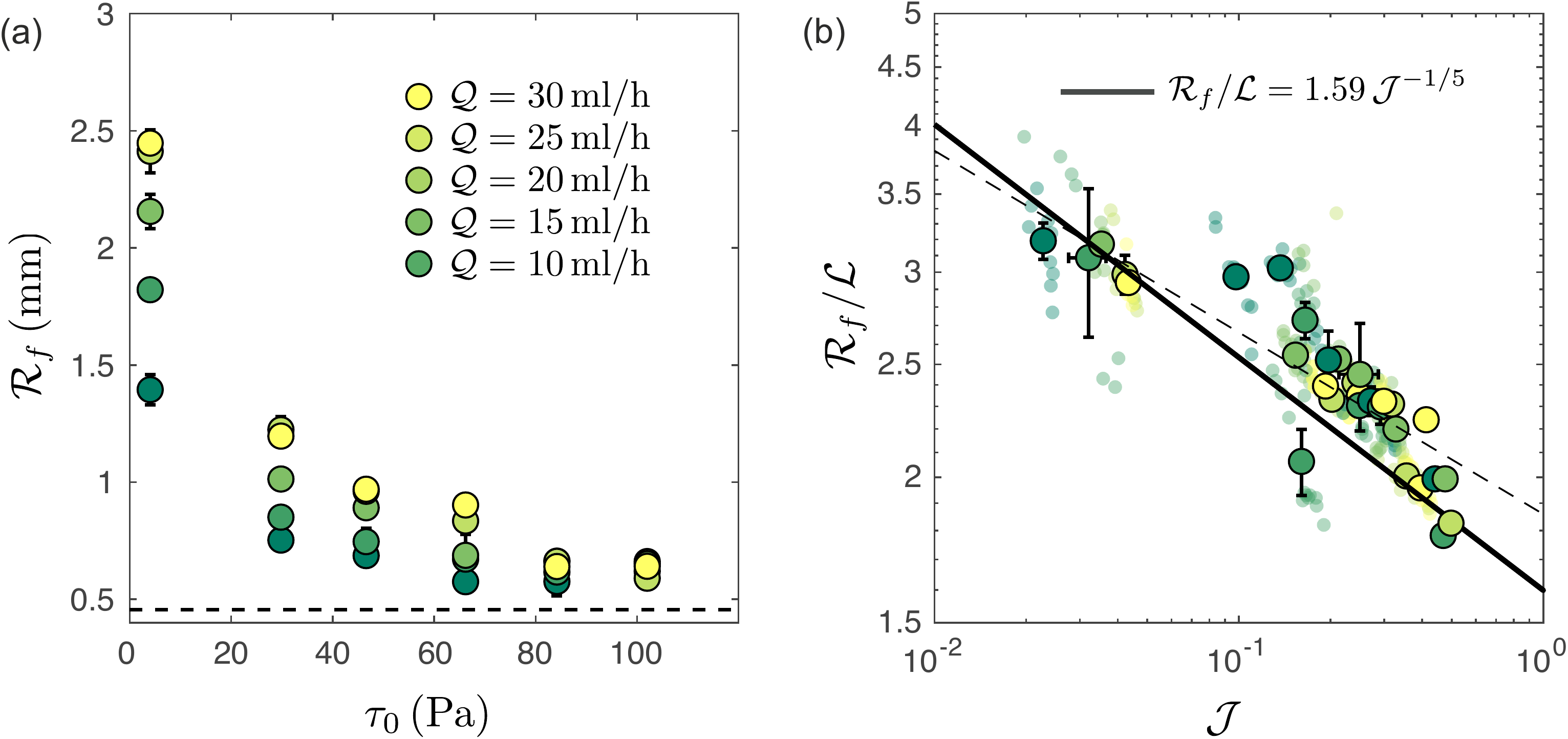}}
  \caption{a) Dimensional final half width of the printed line for different flow rates and yield stress values. The dashed line shows the radius of the nozzle ($\approx 0.46\,$mm).  b) Dimensionless half-width of the lines as a function of the plastocapillarity number $\mathcal{J}$. The experimental repetitions data are plotted as small transparent points and the large circles are the average values with standard deviations as the error bars. Equation~\ref{eq:scaling} is plotted as the solid line. The dashed line is a power-law fit through the experiments, $\mathcal{R}_f / \mathcal{L} = 1.859\, \mathcal{J}^{-0.156}.$ }
\label{fig:exp}
\end{figure}

The experimental results for the final half-width of the filament $\mathcal{R}_f$, for different flow rates $\cal Q$ and yield stresses $\tau_0$, are shown in figure \ref{fig:exp}a. As the yield stress increases, the filaments spread less and thence, the values of $\mathcal{R}_f$ reduce. Meanwhile, for a given yield stress value, a larger flow rate results in a larger deposited volume of the filament and therefore larger $\mathcal{R}_f$. The latter is more pronounced when the yield stress is small. {\color{black} When the values of yield stress is large, the filament is less affected by the surface tension, after the extrusion from the nozzle, Therefore, the final half width of the filament approaches the radius of the nozzle (shown by the horizontal dashed line).}

The experimental conditions were selected such that the filaments' cross-sectional features are small compared to the capillary length scale $l_c =(\sigma/\rho\,g)^{1/2}$. We define the characteristic length scale of the filament as $\mathcal{L} = \sqrt{A}$, where $A(Q,U)$ is the cross-sectional area of the filament obtained by processed OCT images. Depending on the fluid and $Q$, the values of $\mathcal{L}$ vary between 0.2 to 0.9 mm.
The filaments were deposited at relatively low flow rates, $Q = 10-30 \, $ml/h. Given the high apparent viscosity of the liquids and small characteristic length scales of the filaments, the inertial effects are negligible. {\color{black} To this end, one could define a Reynolds number for a Herschel-Bulkley fluid~\citep{jalaal2019viscoplastic} as $Re = \rho \, U^2 /(K\,(U/\mathcal{L})^n + \tau_0)$ which compares the inertial and (apparent) viscous forces and also a Weber number as $We = \rho \, U^2 \, \mathcal{L}/\sigma$ which compares inertial and capillary forces. For the experimental parameters here, $Re \sim We \sim \mathcal{O}(10^{-4})$, showing the negligible inertial effects.} Additionally, the small size of the filaments suggests that the gravitational effects are also {\color{black}relatively small in comparison with the surface tension effects.} The latter could be seen in small values of the Bond number ($Bo=\rho \, g\, \mathcal{L}^2/\sigma$, with, $0.01<Bo<0.13$), which compares the pressure due to the gravitational acceleration ($g$) to capillary pressures. The importance of elastic effects is less obvious, given the large non-linear variation of viscoelastic properties of the liquid under different strains (see figure \ref{fig:rheol}b). 
{\color{black}An estimation for elastic time scale can be made as $\lambda \approx \left(K/G^{\prime}\right)^{1/n}$~\citep{luu2009drop, bird1987dynamics}. For $G^{\prime}$ at the critical strain $\gamma_0$, when stress is close to the yield stress, we measure $\lambda \approx0.1-1\,$s. In our experimental setup, it takes 7-12 seconds for filaments to fully spread on the surface. Hence, the elastic time scales are smaller than the experimental time scales, suggesting the elastic effects might be small. Note that, such a rough comparison of time scales does not necessarily show that elasticity has no effect on the dynamics and final shape of the filament. We will further discuss this in the section \ref{sec:conc}.}

However, a rough comparison of relaxation time scales ($\sim \mathcal{O}(1\, \mathrm{s})$) and the experimental time scales suggest that, at least to a first approximation, the elastic effects are small for the range of experimental conditions here. 
Hence, the final width of the filament is a function of the yield stress $\tau_0$, surface tension coefficient $\sigma$, and the flow rate $Q$. The latter, in our setup, mainly determines the volume of the line ($\sim \cal A$). Therefore, the final width can be found to be a function of a single dimensionless group, namely the plastocapillary number,
\begin{equation}
\mathcal{J} = \frac{\tau_0 \, \mathcal{L}} {\sigma}.
\end{equation}

The number above determines the yield stress ratio to the characteristic capillarity pressure. As the plastocapillary number increases, the plasticity effects become more pronounced compared to the capillary pressure and hence, one can expect a line with a smaller width. {\color{black} In the context of 3D printers, for a millimetric filament with a surface tension close to water, the capillary pressure is $\mathcal{O}(10\,$Pa), meaning inks with a yield stress of $\mathcal{O}(10\,$Pa) or smaller are soft enough to easily spread on the surface. If $\mathcal{J} \gg 1$, one could expect a very stiff filament that sustains the shape after extruding from the nozzle. This means $\mathcal{R}_f/\mathcal{L} \rightarrow 1/\sqrt{\pi}$ when $\mathcal{J}\rightarrow \infty$. }

Figure \ref{fig:exp}b shows the normalized value of the final half-width of filaments ($\mathcal{R}_f/\mathcal{L}$) versus the plastocapillary number for all collected experimental data. The normalized data suggests a power-law relationship between $\mathcal{R}_f/\mathcal{L}$ and $\mathcal{J}$. We attempt to find this relationship in the following.

\subsection{Scaling Law}
Consider a viscoplastic filament that is spreading under capillary actions and comes to a halt due to the yield stress. The final shape of such filament could be approximated by balancing the forces acting on the filament~\citep{jalaal2021spreading}. In this case, the driving force ($F_{\sigma}$) is due to the capillary action, and the resisting force is due to the yield stress ($F_{\tau_0}$), mainly acting on the base of the filament. These forces can be estimated as  
\begin{equation}
F_{\sigma} = \sigma \, \kappa \, \mathcal{H}_f \, b, \quad \mathrm{and,} \quad F_{\tau_0} = \tau_0 \, \mathcal{R}_f \, b,
\label{eq:forces}
\end{equation}
where $\mathcal{H}_f$ is the final height of the filament, $\kappa \approx \mathcal{H}_f/\mathcal{R}_f^2$ is a characteristic curvature, and $b$ is the length of the filament. The cross-sectional area of the filament could be estimated as $A =\mathcal{L}^2\sim\mathcal{H}_f\,\mathcal{R}_f$. Using this, and balancing the forces in equation \ref{eq:forces}, we arrive at
\begin{equation}
\mathcal{R}_f / \mathcal{L} = \Omega \, \mathcal{J}^{-1/5},
\label{eq:scaling}
\end{equation}
where $\Omega$ is a prefactor. Equation \ref{eq:scaling} reveals a relatively weak power-law relationship between the normalized half-width of the filament and the plastocapillary number; as the plastic effects become larger, the filament spreads less. Generally, the values of prefactor $\Omega$ can be a function of gravity, pre-wetted film thickness, elasticity etc. However, for the range of experimental parameters presented here, $\Omega$ can be estimated as a constant, representing the limit of pure plastocapillarity. In the following, we present a thin film approximation of the present problem, in which we also aim to find the value of $\Omega$.

\subsection{Thin Film Approximation}

We apply a lubrication theory for a shallow and inertia-less flow of Bingham viscoplastic fluids (shear index, $n=1$) without gravity effects. Following the previous analyses of viscoplastic lubrication theory \citep{balmforth2019viscoplastic}, and after scaling the length by $\mathcal{L}$, velocity by $\sigma/\mu$, and time by $\mathcal{L}\,\mu/\sigma$, we arrive at the following dimensionless depth-integrated evolution equation:

\begin{equation}
h_t= \frac{1}{6} \left[- h_{xxx} Y^2 \left(3\,h - Y \right) \right]_x, \quad \mathrm{with} \quad Y = \mathrm{max}\left(0, h - \mathcal{J} / \vert h_{xxx} \vert \right),
\label{eq:thinfilm}
\end{equation}
where $h(x,t)$ is the height of the filament and subscripts denote partial derivatives. In equation \ref{eq:thinfilm}, $Y(x,t)$ is the yield surface, corresponding to the position in which the stress is equal to the yield stress. When $\mathcal{J} = 0$, we have $Y=h$ and hence equation \ref{eq:thinfilm} reduces to the well-known thin film equation for Newtonian fluids~\citep{oron1997long,craster2009dynamics}. Equation \ref{eq:thinfilm} represents the planar version of the viscoplastic lubrication theory for droplet spreading~\citep{jalaal2021spreading}. Note that the dynamics of the thin film spreading likely differ for Herschel-Bulkley fluids where $n\neq1$, however, the final shape should still be well approximated in the limit studied here.

\begin{figure}
  \centerline{\includegraphics[width=16cm]{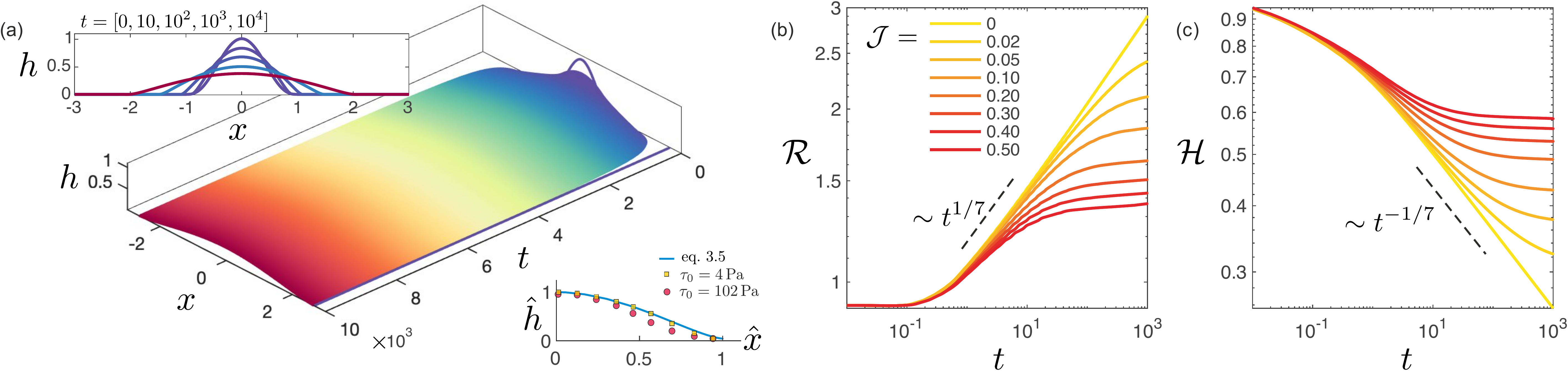}}
  \caption{a) Finite difference solution of equation~\ref{eq:thinfilm}. The 3D dimensional plot is a quasi-representation of a line printing with $\mathcal{J}=0.02$ at $t=10^4$. The solid black line notes the final half-width. The inset on top-left shows the cross section of the line at five different times. The inset on bottom-right compares the theoretical final shape (where $\hat{h}=h/\mathcal{H}_f$ and $\hat{x}=x/\mathcal{R}_f$) with two experimental examples of samples 1 and 6 at $Q=10\,$ml/h, where the width and height are normalized with the theoretical prediction.
  The variation of half-width ($\mathcal{R}$) and maximum height ($\mathcal{H}$)
  of the filament as a function of time for different values of $\mathcal{J}$ are shown in panels b and c, respectively. The dashed black lines show the scaling laws for Newtonian ($\mathcal{J}=0$) limits.}
\label{fig:theo}
\end{figure}

To explore the dynamics of filament spreading, we solve equation \ref{eq:thinfilm} using a finite difference method. In our solution, the length of the domain is large enough such that the filament never arrives at the boundaries. The initial shape of the filament was $h(x,t=0) = \mathrm{max} (0,1-x^2)^3 + h_{\infty}$~\citep{jalaal2021spreading}.
The first and second derivatives of the initial condition are zero at $x=1$ where it connects the interface of the filament to a pre-wetted film of thickness $h_{\infty}$ and has a height of almost unity at the centre with zero derivatives. Note that the pre-wetted film removes a genuine triple contact line and the complexities due to stress singularity~\citep{bonn2009wetting,snoeijer2013moving}. Figure \ref{fig:theo}a shows the numerical solutions of the thin film equation for $\mathcal{J}=0.02$. 
The filament fully yields under surface tension forces at $t=0$ and spreads on the surface. Later in the process, the influence of yield stress becomes more evident as the surface tension forces decrease. The filament eventually stops deforming when stress inside the filament falls below the yield stress ($Y \rightarrow 0$). The variations of half-width $\mathcal{R}(t)$ and maximum height $\mathcal{H}(t)$  of spreading filaments for different plastocapillary numbers are shown in panels b and c of figure~\ref{fig:theo}, respectively. 
In the beginning of the spreading, the dynamics follow the Newtonian spreading~\citep{tanner1979spreading}, for which $\mathcal{R}(t) \sim t^{1/7}$ and $\mathcal{H}(t) \sim t^{-1/7}$, but eventually the curves deviate from the Newtonian limit and approach a constant value that corresponds to the final half-width and height of the filaments. The larger the value of the plastocapillary number is, the sooner the filaments approach a final shape. Also, similar to what has been presented in the experiments, the filaments spread less as the value of the plastocapillary number increases. 

To compare the theoretical results of the final shapes with the experiments, we use the asymptotic limit of equation~\ref{eq:thinfilm} for $Y \rightarrow 0$, at which the flow is ceased. 
Hence, in dimensional form, we have $\sigma \, h\,h_{xxx} = \tau_0$. It is more convenient to re-scale the equation at this limit with final height and half-width, i.e.\ $\hat{h}=h/\mathcal{H}_f$ and $\hat{x}=x/\mathcal{R}_f$. Hence,

\begin{equation}
\hat{h} \, \hat{h}_{\hat{x} \hat{x} \hat{x}} = \Lambda, \quad \mathrm{where,} \quad  \Lambda = \tau_0\,\mathcal{R}_f^3 / \sigma \, \mathcal{H}_f^2. 
\label{eq:final}
\end{equation}

The equation above governs the final shape of the filament with a constant cross-sectional area $A = 2\, \mathcal{H}_f \, \mathcal{R}_f \, \mathcal{I}$, where $\mathcal{I} = \int_0^1 \hat{h} \mathrm{d}\hat{x}$ is \emph{apriori} unknown. From the definition of $\Lambda$ and $A$, we find the same relationship as in~\ref{eq:scaling} with $\Omega = (\Lambda / \mathcal{I}^2)^{1/5}$. We solve equation~\ref{eq:final} as a boundary value problem  (\emph{cf.} the recent work on droplet spreading for the details of numerical method and boundary conditions~\citep{jalaal2021spreading}) {\color{black}to find the final shape} of the filament and $\Lambda \approx 3.53$ and $\mathcal{I} \approx 0.58$, and therefore $\Omega \approx 1.59$. {\color{black}The inset in figure \ref{fig:theo}a compares the obtained theoretical shape with two examples of experimental data with smallest and largest yield stress. While theory predicts the shape of the fluid with small yield stresses very well (red line vs. the OCT image), such comparison is weaker for the fluid with a higher yield stress. This can be seen in both curvature and the values of the final width.} 

A comparison of equation~\ref{eq:scaling} with the obtained $\Omega$ from the asymptotic analysis above is shown in figure~\ref{fig:exp}b. While the theoretical prediction works fine for small values of $\mathcal{J}$, the comparison suffers from discrepancies at higher values. {\color{black}Fitting the data with a power-law function, results in $\mathcal{R}_f / \mathcal{L} = 1.859\, \mathcal{J}^{-0.156}$, showing the theory under-predict the prefactor and over-predict the magnitude of the power law by about 15\%, and 28\%, respectively. The difference between theory and experiments could be due to both theoretical and experimental limitations. We will discuss these limitations in the following section and provide suggestions for future work.}

\section{Conclusion \& Discussion}
\label{sec:conc}
We have experimentally and theoretically studied the printing mechanism of a single filament of a complex fluid with viscoplastic rheology. The study was focused on the plastocapillarity regime, where surface tension is the driving mechanism, and yield stress is the resisting one. {\color{black}The experimental setup offers a simple configuration to study several fundamental aspects of 3D printing of complex fluids. Additionally Optical Coherence Tomography provides opportunity to easily and accurately measure the geometrical features of the filaments and potentially gives insight about the flow inside the filament}. Our primary interest was the final width of the filaments.
Theoretically, we solved the viscoplastic lubrication equations governing the cross-sectional spreading of the filaments. We also found a simple scaling law for the final half-width of the filament $\mathcal{R}_f/\mathcal{L} = 1.59\, \mathcal{J}^{-1/5}$, where the prefactor was obtained via an asymptotic analysis. 
{\color{black}Given the complexity of the problem, and the limitations of the model, the comparison between the experimental and theoretical results shows some discrepancy (see, figure~\ref{fig:exp}).} In experiments, the explicit effects of the pre-wetted film are not fully clear yet. {\color{black}It is expected that a finite thickness of the film affects the spreading~\citep{blackwell2015sticking,sen2020viscoplastic} and possibly slightly increases the final width of the filament~\citep{jalaal2021spreading}.} For the present study, we fixed the values of the pre-wetted film at the thinnest thickness available for our experimental conditions. This allowed us to focus on our main goal, finding the final width of the filament in a capillary-driven regime. One extension of the present work could be a systematic study on the effect of the pre-wetted film thickness on the spreading and final shape of the filament. The experiments were also performed on normal glass surfaces. For droplet spreading on solid surfaces, it has been shown that "apparent slip" could change the dynamics and equilibrium shape of droplets~\citep{jalaal2015slip, martouzet2021dynamic}. Like the pre-wetted film, the \say{slippery} condition of the substrate could result in a larger final width. Hence, these might explain the under-prediction of the theory in figure~\ref{fig:exp}b. Hence, another extension of the work could then be to systematically study the spreading of filaments on rough or chemically treated surfaces to study the effect of \say{slip}.
{\color{black} The theory presented here ignores the elastic effects. In practice, however, materials feature viscoelastic properties below and above the yield stress.
Given that viscoelastic properties can influence the spreading and the final shape of the filament, further investigation is required. A possible extension of our work includes mathematical models with elasto-viscoplastic (EVP) characteristics~\citep{saramito2007new, de2011thixotropic, dimitriou2014comprehensive, saramito2017progress}. Such models could potentially be implemented in thin-film limit or numerical simulations which solve for two phases with moving boundaries~\citep{izbassarov2020dynamics, sanjay2021bubble}. Using EVP models, one should be able to explicitly study different elastic effects.}
Also note that the presented theory assumes the cross sections of the filaments have a large aspect ratio (they are shallow). While valid for most of the experimental parameters, for the largest yield stress, this aspect ratio is around 6. The theory could be improved in non-asymptotic fashions, by retaining the full surface curvature or using numerical simulations to go beyond the lubrication limit.

Our work has direct applications in the coating and 3D printing industries, where spreading and printing filaments of viscoplastic fluids play a key role. However, by studying a single filament of soft yield stress materials, we have not addressed a number of relevant questions in the broad context of 3D printing. Further extensions of the present study include the spreading and \textit{welding} of multiple filaments next to each other, or on top of each other. Additionally, the regime of very large plastocapillary numbers still requires more investigation, where the deformations due to surface tension are much smaller and localized. Such investigations are required to fully address the printability of ink in relationship to rheology and fluid mechanics of 3D printing. Moreover, a more systematic study of the flow of viscoplastic fluids near the nozzle, its effects on the filament shapes, printing on dry surfaces when contact angle dynamics play and important role, and printing more complicated line geometries like corners~\citep{friedrich2020corner} are still required. Lastly, the present study mostly focuses on the final state of the filaments. More systematic experimental and theoretical studies on the spreading of filaments with complex rheology are yet to be done to reveal the dynamics of spreading. This is of great importance when the material itself features rheological properties such as shear-thinning, elasticity, aging, and thixotropy~\citep{chen2018novel,corker20193d, sen2021thixotropy,sen2020viscoplastic}, factors that were not studied in the present work.

The OCT easily furnishes cross-sectional images of the filament at small scales and the points it merges to the pre-wetted film. Further improvement of the system can result in simultaneous measurement of the dynamics of the spreading, rheological features of the fluid, and the flow field inside the filaments~\citep{manukyan2013imaging, trantum2014biosensor,edwards2018density,jalaal2018gel}.\\

\textbf{Acknowledgement}
We are grateful to N.J. Balmforth for helping with the viscoplastic lubrication theory. We would like to thank U. Sen, C. Seyfert, V. Sanjay, and D. Giesen for useful discussions. 

\textbf{Declaration of interests:} The authors report no conflict of interest.







\printbibliography


\end{document}